\documentclass{elsart}

\usepackage{amssymb}
\newcommand{\be}{\begin{equation}}
\newcommand{\ee}{\end{equation}}
\newcommand{\bea}{\begin{eqnarray}}
\newcommand{\nn}{\nonumber}
\newcommand{\eea}{\end{eqnarray}}

\begin{document}

\begin{frontmatter}


\title{Curvature scalar instability in $f(R)$ gravity}
\author{Thomas P.~Sotiriou}
\ead{sotiriou@sissa.it}
\address{SISSA-International School for Advanced Studies, via Beirut 2-4, 34014, Trieste, Italy and INFN, Sezione di Trieste, Italy}





\begin{abstract}
An instability in the presence of matter in theories of gravity which include a $1/R$ correction in the gravitational action has been found by Dolgov and Kawasaki. In the present paper this instability is discussed for $f(R)$ gravity in general. We focus on the Palatini formalism of the theory and it is shown that no such instability occurs in this version of $f(R)$ gravity. The reasons for the appearance of the instability in the metric but not in the Palatini formalism are fully investigated.  
\end{abstract}

\end{frontmatter}

Modified gravity has been attracting increased interest recently, since it constitutes an interesting option for explaining the nature of dark energy, which according to observation should account for about 70\% of the energy density of the universe and be responsible for its current accelerated expansion \cite{Spergel:2006hy}. What is more, our current candidates for a more fundamental theory which will be able to describe gravity in a high energy regime do predict corrections in the low-energy effective action for gravity, with respect to the well-known Einstein-Hilbert action \cite{Buch}.

The simplest form of such a generalized action seems to be
\be
\label{action}
S=\frac{1}{2\kappa^2}\int d^4 x \sqrt{-g} f(R) +S_M(g_{\mu\nu},\psi).
\ee
where $\kappa^2=8\pi G$, $S_M$ is the matter action, $\psi$ collectively denotes the matter fields and $f$ is a general function of the Ricci scalar $R$.
The field equations of General Relativity (GR) can be derived from this action by varying with respect to the metric and applying the stationary action principle, once we assume that $f(R)=R$. In the most general case one derives the following field equation \cite{buchdahl}:
\be
\label{metf}
f'(R)R_{\mu\nu}-\frac{1}{2}f(R)g_{\mu\nu}-\nabla_\mu\nabla_\nu f'(R)+g_{\mu\nu}\Box f'=\kappa^2 T_{\mu\nu},
\ee
where $\nabla$ is the covariant derivative defined with the Levi-Civita connection of the metric, $T_{\mu\nu}\equiv\frac{-2}{\sqrt{-g}}\frac{\delta S_M}{\delta g^{\mu\nu}}$, $\Box\equiv \nabla^\mu\nabla_\mu$ and $A'(x)\equiv \partial A(x)/\partial x$.

Starting from the the same action, one can think of treating the metric and the connection as completely independent entities. In this case $R$ will be replaced everywhere with $\tilde{R}=g^{\mu\nu}R_{\mu\nu}(\Gamma)$, where $R_{\mu\nu}(\Gamma)$ is the Ricci tensor constructed in the usual way with the independent connection $\Gamma^{\lambda}_{\phantom{a}\mu\nu}$. Assuming that this connection is symmetric (torsionless), the independent variation leads to the field equations \cite{ffr,sotlib}
\bea
\label{field1sym}
& &f'(\tilde{R}) \tilde{R}_{(\mu\nu)}-\frac{1}{2}f(\tilde{R})g_{\mu\nu}=\kappa^2 T_{\mu\nu},\\
\label{field2sym}
& &\frac{1}{\sqrt{-g}}\bigg[-\tilde{\nabla}_\lambda\left(\sqrt{-g}f'(\tilde{R})g^{\mu\nu}\right)+\tilde{\nabla}_\sigma\left(\sqrt{-g}f'(\tilde{R})g^{\sigma(\mu}\right)\delta^{\nu)}_\lambda\bigg]=0,
\eea
where $\tilde{\nabla}$ is the covariant derivative defined with the independent connection. This constitutes what is often called in the literature, the Palatini formalism. Notice that this approach assumes that the matter action does not depend on the connection, and is therefore not suitable for all types of matter, such as fermions for example, unless one imposes the extra condition that matter only couples to the Levi-Civita connection of the metric \cite{sotlib,Sotiriou:2006hs}. What is more, it would be natural to allow the connection to be non-symmetric if we wanted to consider fermions. There exists a natural generalization of the Palatini formalism, the metric-affine formalism, which does indeed have these characteristics and is therefore much more appropriate in the present context \cite{sotlib,hehl}. In the metric-affine formalism there is a matter term on the left hand side of eq.~(\ref{field2sym}) coming from the variation of the matter action with respect to the connection. However, in the present paper we will restrict ourselves to the Palatini formalism since this study will focus on instabilities that appear in the weak field and therefore we will not need to consider fermions. In this case the Palatini and the metric-affine formalisms actually coincide, so our results will apply for both cases.

$f(R)$ gravity, both in the metric and the Palatini formalism, has been widely used to address the dark energy problem and it has been shown that models in which $f(R)$ includes terms inversely proportional to $R$ (or $\tilde{R}$) can account for the late time accelerated expansion of the universe \cite{cosm1,cdtt}. Note that positive powers of the curvature can provide a mechanism for early time inflation as well \cite{cosm2}. However such models are not problem free. For example, it is still a matter of investigation whether there can be a smooth passage from one cosmological era to the other and whether this can give a cosmological evolution including all eras \cite{amendola}. There is also still debate on which models have the correct weak field limit or even which is the correct way to obtain that \cite{PPN}. However, our concern in this paper is to study an instability discovered by Dolgov and Kawasaki \cite{DolgovKawasaki} in the presence of matter for the specific model where $f(R)=R-\mu^4/R$, with $\mu$ being a constant \cite{cdtt}. This instability is not just a special characteristic of this model but occurs in a more general class \cite{valerio1}.

Let us briefly review the results of \cite{DolgovKawasaki,valerio1}. By contracting eq.~(\ref{metf}) one gets
\be
\label{inst}
3\Box f'(R)+f'(R)R-2f(R)=\kappa^2 T,
\ee
where $T=g^{\mu\nu}T_{\mu\nu}$. Following \cite{valerio1} we can write $f(R)=R+\epsilon \varphi(R)$, where $\epsilon$ is a constant.
 If we consider a small region in a weak field regime within matter we can assume that 
$g_{ab}=\eta_{ab}+h_{ab}$ and $R=-\kappa^2\, T +R_1$,
where $\eta_{ab}$ is the Minkowski metric and $\left| R_1/\kappa^2\, T\right| \ll 1$. In this approximation, and to first order in  $R_1$, eq.~(\ref{inst}) gives
\begin{eqnarray}
\label{veq}
\ddot{R}_1 -\nabla^2 R_1 &-&\frac{2\kappa^2\, \varphi '''}{\varphi ''}\, (\dot{T}\dot{R}_1-\vec{\nabla}T \cdot \vec{\nabla}R_1)\nn\\&+&\frac{1}{3\varphi ''} \left( \frac{1}{\epsilon}-\varphi' \right) R_1=\kappa^2 \, \ddot{T}-\kappa^2 \nabla^2 T -\, \frac{\left(\kappa^2\, T\varphi '+\varphi \right)}{3\varphi ''},
\end{eqnarray}
where an over-dot denotes differentiation with respect to time, whereas $\vec{\nabla}$ and $\nabla^2$ denote the gradient and Laplacian operators respectively in Euclidean  
three-dimensional space.

The instability occurs if $\varphi ''=f''(R)<0$ and if $\epsilon$ is very small, since the coefficient of the last term on the left hand side of eq.~(\ref{veq}) is the square of an effective mass. As already mentioned in \cite{valerio1} it can be considered as an instability in the gravity sector. Because of this, and since it appears in the equations governing the dynamics of the curvature scalar, we refer to it as ``curvature scalar instability''. Theories with $f''(R)>0$ will be stable irrespective of the value of $\epsilon$. However, for several models that lead to the desired cosmological dynamics at late times, $\epsilon$ is indeed very small and $f''(R)$ is indeed negative. A typical example is the model of \cite{cdtt}, where $\varphi(R)=-\mu^4/R$, with $\mu\sim 10^{-33}$eV, and the time scale for the instability to occur is of the order of $10^{-26}$ s \cite{DolgovKawasaki}.

All of the above is with reference to the metric formalism. Let us now consider the Palatini formalism. We will argue that such an instability cannot occur in this case irrespective of the form of the Lagrangian. Contracting eq.~(\ref{field1sym}) gives
\be
\label{stru}
f'(\tilde{R})\tilde{R}-2f(\tilde{R})=\kappa^2 T.
\ee
Note that $\tilde{R}$ is not the Ricci scalar of the metric. Eq.~(\ref{field2sym}) can actually take the form $\tilde{\nabla}_\lambda h_{\mu\nu}=0$ if one defines $h_{\mu\nu}$ as $h_{\mu\nu}=f'(\tilde{R})g_{\mu\nu}$, which implies that the connections are the Christoffel symbols of this metric \cite{ffr,sotlib}. We can then express $R$ in terms $\tilde{R}$:
\be
\label{R}
R=\tilde{R}-\frac{3}{2[f'(\tilde{R})]^2}\nabla_\mu f'(\tilde{R}) \nabla^\mu  f'(\tilde{R})+\frac{3}{ f'(\tilde{R})}\Box  f'(\tilde{R}).
\ee
Now notice that eq.~(\ref{stru}) is an algebraic equation in $\tilde{R}$ for a given $f(\tilde{R})$ which will have solutions of the form 
$\tilde{R}=\theta(T)$,
where $\theta$ is some function. We are not going to examine cases in which eq.~(\ref{stru}) has no solutions or is identically satisfied ($f(\tilde{R})\propto \tilde{R}^2$), since these do not constitute viable choices for a low-energy gravitational theory \cite{ffr,sotlib}. 

We can now write eqs.~(\ref{R}) as
\bea
\label{matter}
R=\theta(T)-\frac{3}{2[f'(\theta(T))]^2}\nabla_\mu f'(\theta(T)) \nabla^\mu  f'(\theta(T))+\frac{3}{ f'(\theta(T))}\Box  f'(\theta(T)),\nn\\
\eea
or alternatively
$R=F(T)$,
where $F(T)$ is a function of $T$. This clearly demonstrates that the Ricci scalar of the metric can be expressed directly as a function of the trace of the stress-energy tensor. In fact, eq.~(\ref{matter}) is a straightforward generalization of the contracted Einstein equation, $R=-\kappa^2 T$. From the form of eq.~(\ref{matter}), it is clear that no instability can occur in this case, since $R$ carries no dynamics in eq.~(\ref{matter}), unlike eq.~(\ref{inst}).

Let us now analyze where this difference between the two formalisms stems from. By generalizing the Lagrangian from $R$ or $\tilde{R}$ one adds inevitably a scalar degree of freedom \cite{Sotiriou:2006hs}. However this degree of freedom seems to be of a different nature in the two versions of the theory. In the metric version it is dynamical, and therefore care should be taken to ensure stability, whereas in the Palatini version, it is non dynamical. This is related to the fact that the Palatini formalism leads to second order equations as opposed to the metric formalism which leads to fourth order field equations in the metric (see \cite{sotlib} for a discussion), but it can also be seen easily by using the equivalence of $f(R)$ gravity and scalar-tensor theory \cite{equiv,Sotiriou:2006hs}. 

Introducing an auxiliary field $\chi$, we can write the action for the metric formalism as \cite{equiv,Sotiriou:2006hs}
\be
\label{metactionH}
S=\frac{1}{2\kappa^2}\int d^4 x \sqrt{-g} \left(f(\chi)+f'(\chi)(R-\chi)\right) +S_M(g_{\mu\nu},\psi).
\ee
Variation with respect to $\chi$ leads to the equation $\chi=R$ if $f''(\chi)\neq 0$, which reproduces action (\ref{action}).
Redefining the field $\chi$ by $\Phi=f'(\chi)$ and setting
$V(\Phi)=\chi(\Phi)\Phi-f(\chi(\Phi))$
 the action takes the form
\be
\label{metactionH2}
S_{}=\frac{1}{2\kappa^2}\int d^4 x \sqrt{-g} \left(\Phi R-V(\Phi)\right) +S_M(g_{\mu\nu},\psi),
\ee
which is the action of a scalar-tensor theory with $\omega_0=0$. In the Palatini formalism the action will be formally the same apart from the fact the $R$ will become $\tilde{R}$ and $\Phi$ will become $\tilde{\Phi}$,
but in this case this will not be a scalar-tensor theory with  $\omega_0=0$ since $\tilde{R}$ is not the Ricci scalar of the metric \cite{Sotiriou:2006hs}. However if we use eq.~(\ref{R}), and with $\tilde{\Phi}=f'(\tilde{R})$, we get
\be
\label{palactionH2d0}
S_{pal}=\frac{1}{2\kappa^2}\int d^4 x \sqrt{-g} \left(\tilde{\Phi} R+\frac{3}{2\tilde{\Phi}}\partial_\mu \tilde{\Phi} \partial^\mu \tilde{\Phi}-V(\tilde{\Phi})\right) +S_M(g_{\mu\nu}, \psi),
\ee
which is indeed a scalar tensor theory, but with $\omega_0=-3/2$.

The field equation of the scalar field in scalar-tensor theory is 
\be
\label{bdf2}
(2\omega_0+3) \Box \phi= \kappa^2 T+\phi V'-2V.
\ee
Note that $\Phi$ and $\tilde{\Phi}$ are the extra degrees of freedom of $f(R)$ gravity, with respect to General Relativity, in the metric and in the Palatini formalism respectively. Using eq.~(\ref{bdf2}) it is obvious that they satisfy the field equations
\bea
3\Box \Phi+2V(\Phi)-\Phi V'(\Phi)&=& \kappa^2 T,\\
2V(\tilde{\Phi})-\tilde{\Phi}V(\tilde{\Phi}) &=& \kappa^2 T.
\eea 
This demonstrates that $\Phi$ is indeed dynamical, whereas $\tilde{\Phi}$ is not, as mentioned above. At this point, it is worth mentioning that one should not be misled into judging the dynamics of a non-minimally coupled field by the presence or absence of a kinetic term in the action. There are no kinetic terms for $\Phi$ in action (\ref{metactionH2}) but it is still dynamical. Exactly, the opposite holds for $\tilde{\Phi}$. The reason for this is that both fields are coupled not only to the metric, but also to its derivatives. Therefore, when varying the action with respect to the metric and then integrating by parts in order to ``free'' $\delta g^{\mu\nu}$, terms including derivatives of the scalar field are bound to appear. Therefore, even though there are no apparent kinetic terms for $\Phi$ in the action, there will be kinetic terms in the field equations. For $\tilde{\Phi}$, $\omega_0=-3/2$ and this is the remarkable case where these kinetic terms exactly cancel out the ones coming from the kinetic part of the action.

It has been shown that the curvature scalar instability discovered by Dolgov and Kawasaki for metric $f(R)$ gravity is not present in the Palatini formalism, irrespective of the functional form of $f$. This is due to the fact that, even though the generalization of the Lagrangian from $R$ or $\tilde{R}$ to $f(R)$ or $f(\tilde{R})$ always corresponds to intoducing an additional scalar degree of freedom in the theory, this degree of freedom is dynamical in the metric formalism , whereas in the Palatini case it is not. This instability does not occur in Palatini $f(R)$ gravity but other types of instabilities might well be present. For example, judging from the form of eq.~(\ref{matter}) it is not difficult to imagine that specific forms of $f$ could lead to a blow-up of the scalar-curvature for small density perturbations around a stable matter configuration, which are, of course, of a different nature. This issue also seems to be directly related to the weak field limit of the theory. We hope to address it in future work.

{\em Acknowledgements:}
The author would like to thank Valerio Faraoni for enlightening discussions and comments and John Miller and Stefano Liberati for a critical read of this manuscript and constructive suggestions.




\end{document}